\documentclass[journal=jacsat,manuscript=article]{achemso}

\usepackage[version=3]{mhchem} 
\usepackage{graphicx} 

\usepackage[english]{babel}
\usepackage{amsmath}
\usepackage{amsfonts}
\usepackage{bm}
\usepackage[utf8]{inputenc}
\usepackage{soul} 
\usepackage{xcolor}
\usepackage{listings}
\usepackage{soul}
\usepackage[capitalise]{cleveref}
\crefname{equation}{Eq.}{Eqs.}
\usepackage{siunitx}
\DeclareSIUnit{\mbar}{mbar}
\DeclareSIUnit{\litre}{l}
\DeclareSIUnit{\liter}{l}

\newcommand{\mr}[1]{\ensuremath{\mathrm{#1}}}

\newcommand{\vmzi}{\ensuremath{\langle \bar{\bm z}_0 \rangle}}
\author{Martin Ducha\v{n}}
\affiliation[]
{Institute of Scientific Instruments of the Czech Academy of Sciences, Kr\'{a}lovopolsk\'{a} 147, 612~64 Brno, Czech Republic}
\alsoaffiliation[]
{Department of Theoretical Physics and Astrophysics, Faculty of Science, Masaryk University, Kotl\'{a}\v{r}sk\'a 267/2, 611~37 Brno, Czech Republic}

\author{Alexandr Jon\'a\v{s}}
\affiliation[]
{Institute of Scientific Instruments of the Czech Academy of Sciences, Kr\'{a}lovopolsk\'{a} 147, 612~64 Brno, Czech Republic}

\author{Radim Filip}
\affiliation[]
{Department of Optics, Palack\' y University, 17. listopadu 1192/12,  771~46 Olomouc, Czech Republic}

\author{Jan Je\v{z}ek}
\affiliation[]
{Institute of Scientific Instruments of the Czech Academy of Sciences, Kr\'{a}lovopolsk\'{a} 147, 612~64 Brno, Czech Republic}

\author{Petr J\'akl}
\affiliation[]
{Institute of Scientific Instruments of the Czech Academy of Sciences, Kr\'{a}lovopolsk\'{a} 147, 612~64 Brno, Czech Republic}

\author{Pavel~Zem\'anek}
\email{zemanek@isibrno.cz}
\affiliation[]
{Institute of Scientific Instruments of the Czech Academy of Sciences, Kr\'{a}lovopolsk\'{a} 147, 612~64 Brno, Czech Republic}

\author{Martin \v{S}iler}
\email{siler@isibrno.cz}
\affiliation[]
{Institute of Scientific Instruments of the Czech Academy of Sciences, Kr\'{a}lovopolsk\'{a} 147, 612~64 Brno, Czech Republic}

\title[An \textsf{achemso} demo]
  {Fermat's Spiral-Based Characterization of Squeezed Nonlinear Motional States of
Levitated Nanoparticle}

\abbreviations{IR,NMR,UV}
\keywords{non-Gaussian states, motional squeezing, Fermat's spiral, optical tweezers, levitated nanoparticle}

\begin{document}

\begin{tocentry}
\includegraphics{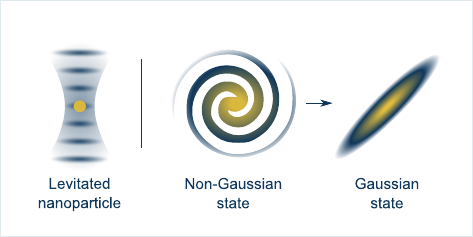}





\end{tocentry}

\begin{abstract}
  Controlling the state of motion of optically levitated nanoparticles is crucial for the advancement of precision sensing, fundamental tests of physics, and the development of hybrid classical-quantum technologies. Experimentally, such control can be achieved by pulsed modifications of the optical potential confining the nanoparticle. 
  Most frequently, the applied potential pulses are parabolic in nanoparticle position, and they   
  expand/squeeze or displace the initial Gaussian state of motion to a modified Gaussian state. 
  The time-dependent mean values and covariance matrix of the phase-space variables can fully characterize such a state. 
  However, quasi-parabolic optical potentials with added weak Duffing-type nonlinearity, encountered in real-world experiments, can generally distort the state of motion to a non-Gaussian one, for which the description based solely on the mean values and covariance matrix fails. Here, we introduce a nonlinear transformation of the phase-space coordinates using the concept of Fermat’s spiral, which effectively removes the state distortion induced by the Duffing-type nonlinearity and enables characterization of the state of motion by the standard Gaussian-state metrics. Comparisons of the experimental data with theoretical models 
  show that the proposed coordinate transformation can recover the ideal behavior of a harmonic oscillator even after extended evolution of the system in the nonlinear potential. The presented scheme enables the separation of the effects of the applied state manipulation, the system's gradual thermalization, and the nonlinearity of the confinement on the experimentally observed dynamics of the system, thereby facilitating the design of advanced protocols for levitated optomechanics.
\end{abstract}


\section{Introduction}
    Microscale stochastic systems, whose behavior is governed by the interplay between deterministic dynamical laws and random forcing resulting from the contact with an ambient thermal environment, are found in virtually all domains of natural sciences ranging from physics through chemistry to biology~\cite{chandrasekhar_stochastic_1943,hanggi_artificial_2009}.
    In an analogy with the quantum-mechanical formalism~\cite{schleich_quantum_2001,rosiek_quadrature_2024}, the randomly fluctuating mechanical state of such microsystems can be quantitatively described using the time-dependent probability density function (PDF), $P(z,v,t)$, in the phase space with the conjugate canonical coordinates of position $z$ and velocity $v$. For a system with linear dynamics, $P(z,v,t)$ is Gaussian-distributed with respect to $(z,v)$ and, consequently, can be fully characterized by the mean values and covariances of $(z,v)$ that generally vary with time $t$~\cite{siler_bayesian_2023}.
    
    The state of a linear dynamical system can be manipulated by subjecting the system to well defined external forces~\cite{rashid_experimental_2016,hebestreit_sensing_2018,bonvin_state_2023,muffato_generation_2024,duchan_nanomechanical_2025}. If these forces are conservative and at most linear in position (i.e., can be described by a potential at most quadratic in position), these manipulations do not change the original Gaussian character of the PDF describing the state of the system. 
    Controlled engineering and manipulation of Gaussian states is widely used in quantum mechanics~\cite{kamba_revealing_2023,rossi_quantum_2024}; its classical counterpart then allows experimental testing of the fundamental concepts (e.g., state squeezing) and specific manipulation protocols under straightforwardly accessible laboratory conditions.
    
    Nanoparticles optically levitated in vacuum represent a versatile experimental platform for testing the various state-manipulation protocols, offering a large degree of control over the characteristic parameters of the forces and of the thermal environment that govern the stochastic dynamics of the system~\cite{gonzalez-ballestero_levitodynamics_2021,millen_optomechanics_2020,winstone_optomechanics_2024,jin_towards_2024}. 
    The state-manipulation protocols, realized by controlled changes of the optical trapping potential~\cite{rashid_experimental_2016,hebestreit_sensing_2018,bonvin_state_2023,kamba_revealing_2023,muffato_generation_2024,duchan_nanomechanical_2025}, can be used for parametric studies of physical concepts at the classical/quantum boundary (e.g., manifestations of non-Hermiticity and entanglement~\cite{arita_optical_2018,svak_stochastic_2021,arita_all-optical_2022,liska_cold_2023,liska_pt-like_2024,reisenbauer_non-hermitian_2024,rieser_tunable_2022}) and for implementation of sensitive methods of detection of weak forces and torques~\cite{moore_searching_2021,monteiro_force_2020,li_collective-motion-enhanced_2023}.
    
    The nanoparticles confined near the minimum of the optical trapping potential that corresponds to the maximum of optical intensity of the incident light are usually modeled as linear dynamical systems characterized by a constant value of the trap stiffness. However, for larger displacements of the trapped nanoparticle from the potential minimum, real implementations generally deviate from this simplifying assumption~\cite{jones_optical_2015,gieseler_optical_2021}. 
    Nonlinearity of the trapping potential inevitably destroys the Gaussian character of the particle's state that can no longer be characterized solely in terms of mean values and covariances of the phase-space variables $(z,v)$~\cite{duchan_nanomechanical_2025,setter_characterization_2019}. Consequently, the linear protocols for the manipulation of Gaussian states and their various applications break down.
    
    For the frequently used optical trapping schemes based on a single focused laser beam or on a standing wave, a realistic trapping potential can be well represented by the Duffing model, with the effective trap stiffness weakening for larger displacements from the potential minimum due to the decay of the optical intensity of the trapping light from its maximum at the trapping location. 
    Confinement in the quasi-parabolic Duffing potential with a weak quartic term~\cite{strogatz_nonlinear_2007,kovacic_duffing_2011} leads to oscillations of the trapped particle with the frequency dependent on the oscillation amplitude~\cite{flajsmanova_using_2020}. As demonstrated below, this amplitude-dependent oscillation frequency then leads to a time-dependent, spiral-shaped distortion of the phase-space PDF describing the evolution of the state of the particle. Formation of such spiraling patterns is ubiquitous among a diverse group of nonlinear dynamical systems ranging from the astronomic~\cite{dobbs_dawes_2014} to the quantum~\cite{abo-shaeer_observation_2001,nicolin_faraday_2007} scale.
    
    In this paper, we experimentally and theoretically investigate the nonlinear deformations of an initially Gaussian PDF characterizing the motion of an optically levitated nanoparticle, induced by the Duffing nonlinearity. To retrieve the validity of the complete characterization of the state of motion using the standard linear dynamical metrics of the mean values and covariance matrix of the phase-space coordinates, we introduce nonlinear transformations of these coordinates that are consistent with the observed spiral-like shape of the PDF support. 
    Our analysis shows that the coordinate transformation based on Fermat's spiral~\cite{lockwood_book_2007}, which correctly reflects the dynamical laws governing the studied system, effectively restores the Gaussian character of the nanoparticle's state following an extended confinement in the Duffing potential. Thus, the presented approach enables the analysis of squeezed nonlinear stochastic states beyond the Gaussian approximation.

\section{State description and evolution in phase space}
\subsection{State description}
    Let us assume first that the nanoparticle of mass $m$ oscillates in a pure parabolic potential of stiffness $\kappa$ with a characteristic angular frequency $\Omega_0=\sqrt{\kappa/m}$.
    Due to the random interaction with surrounding molecules or photons, its location in the phase space becomes uncertain. It can be characterized by a time-dependent PDF denoted as $P(z,v, t)$. 
    One may rescale the phase space coordinate system $(z,v)$ to dimensionless coordinates in various ways. Here we used $(\bar z, \bar v)$ normalized to equilibrium 
    standard deviation of the particle in the trap or to the particle's mean thermal velocity\cite{li_non-hermitian_2024}:
    \begin{equation}
        \bar z = z \sqrt{\frac{m\Omega_0^2}{k_\mr{B}T}}, \,\,\,
        \label{eq:normz} 
        \bar v = v \sqrt{\frac{m}{k_\mr{B}T}}
    \end{equation}
    where $k_\mr{B}$ is the Boltzmann constant, and $T$ is the thermodynamic temperature of the ambient gas. 

    The PDF of particle phase space occurrence at a given time of a Gaussian state can be given as \cite{chandrasekhar_stochastic_1943}
    \begin{equation}
        P(\bar z, \bar v, t) = \frac{1}{2\pi \sqrt{\mid \bm{\bar \Theta} \mid}} \exp\left\{-\frac{1}{2} \left(\bm{\bar z} -\langle\bm{\bar \bm z_0}\rangle\right)^T \bm{\bar \Theta}^{-1} \left(\bm{\bar z} - \langle\bm{\bar \bm z_0}\rangle\right)\right\},
        \label{eq:GaussPDF}
    \end{equation}
    where 
    \begin{equation}
        \bm{\bar z} = \left( \begin{array}{c}\bar z\\ \bar v \end{array} \right), \quad 
        \langle\bm{\bar \bm z_0}\rangle = \left( \begin{array}{c} \langle \bar z_0(t) \rangle\\ \langle \bar v_0(t) \rangle \end{array} \right),
        \quad \bm{\bar \Theta} = \left( \begin{array}{cc}\bar \theta_{zz}(t)&\bar \theta_{zv}(t)\\ \bar \theta_{zv}(t)& \bar \theta_{vv}(t) \end{array} \right)
        \label{eq:GaussCM}
    \end{equation}
    denote the phase space location vector, the time-dependent vector of probability density mean values, and the time-dependent covariance matrix, respectively, and $\mid \bm{\bar \Theta} \mid$ is the determinant of $\bm{\bar \Theta}$. 
    
\begin{figure}[H]
        \centering
        \includegraphics[width=1\linewidth]{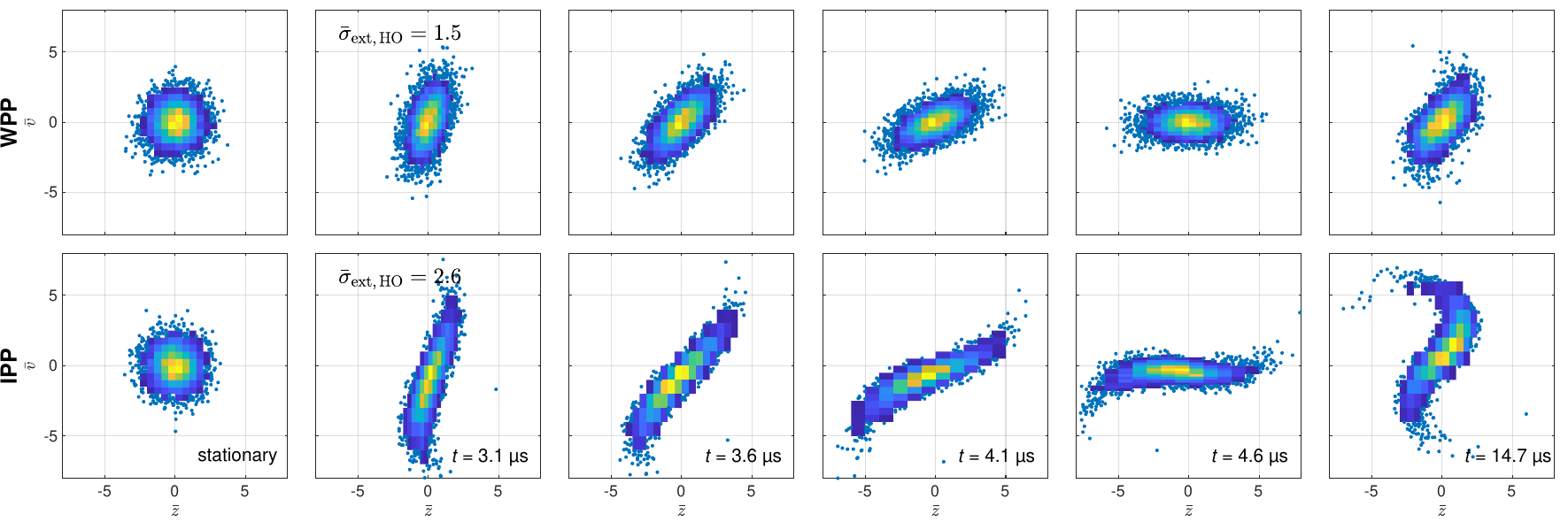}
        \caption{\textbf{Evolution of the probability density function (PDF) of nanoparticle location in phase space}. Colored maps show the probability density of detecting the nanoparticle at a given element of phase space at several times from the initial Gaussian state. Individual dots denote less-frequently detected events. The left column shows the thermal state from which the the initial state for the quasi-parabolic potential is engineered by a weak parabolic (WPP, top row, pulse duration $\tau= \SI{1}{\micro\second}$) or inverted parabolic (IPP, $2\times$ longer pulse of duration $\tau= \SI{2}{\micro\second}$, bottom row) potential pulse. The first measured nanoparticle motional state (at $t=\SI{3.1}{\micro\second}$) is drawn in the second column. Its time evolution in the quasi-parabolic potential follows in the subsequent columns. 
        Due to the longer IPP pulse, the corresponding PDF extension is larger. Its outer parts reach deeper into regions of the quartic nonlinearity and consequently, due to \cref{eq:varphi}, they rotate more slowly and lag behind the central parts. The parameter $\bar \sigma_{\mathrm{ext,\,HO}}$ represents the theoretical state extension for an ideal harmonic oscillator. More experimental details are described in SI.}
        \label{fig:evolution}
    \end{figure}

    By setting $\bar \mu_z \equiv \bar \mu_v \equiv \bar \theta_{zv} \equiv \bar \theta_{vz} \equiv 0$ and $ \bar{\theta}_{zz} \equiv \bar{\theta}_{vv} \equiv 1$, the PDF describes the state of thermal equilibrium corresponding to the ambient temperature $T$ in \cref{eq:normz}. An example of such motional states obtained experimentally can be seen in the left column in \cref{fig:evolution}.
    
    A specific Gaussian state at $t=0$, defined by its mean values $\vmzi$ and covariance matrix  $\mathbf \Theta_0$, can be engineered by an appropriate sequence of parabolic potential pulses \cite{duchan_nanomechanical_2025} from the thermal equilibrium state corresponding to the temperature $T$. We investigated the following two types of such pulses (see \cref{fig:protocol})
    \begin{itemize}
        \item parabolic -- weak parabolic (WPP)
        \item  parabolic -- displaced inverted parabolic (IPP).
    \end{itemize}
    Development of various initial states in the quasi-parabolic potential at $t>0$ will be the subject of our investigation below. 
    
    \subsection{Time evolution in parabolic potential}
    The initial Gaussian state, defined by initial mean values $\vmzi$ and covariance matrix  $\mathbf \Theta_0$, evolves into a new Gaussian state in a linear regime defined by an evolution matrix $\mathbf {U}(t)$, $\bm f(t)$\cite{duchan_nanomechanical_2025}, corresponding to an appropriate parabolic potential, and an additional external constant force $F_\mr{c}$, causing displacement of mean values \cite{duchan_nanomechanical_2025}: 
    \begin{eqnarray}
     \langle \bar{\bm z}(t) \rangle &=& \mathbf U(t) \vmzi + \bm f(t)   \bar{F}_\mr{c} \\
     \mathbf {\bar \Theta}(t) &=& \mathbf {\bar \Theta_{\mr{f}}}(t) + \mathbf {U}(t) \mathbf {\bar \Theta_0 U}(t)^T     \label{eq:Thetadef}
    \end{eqnarray}
    where $\mathbf {\bar \Theta_\mr{f}}$ describes the temperature-dependent diffusive evolution. 
    The particular forms of the matrices for linear regimes can be found in the SI of \cite{duchan_nanomechanical_2025}.
    In a simplified way, the PDF in the form of an elongated two-dimensional Gaussian in the normalized phase space coordinates rotates clockwise in time with an angular frequency corresponding to the oscillation frequency $\omega$ of the damped harmonic oscillator 
    \begin{equation}
     \omega = \left(\Omega_0^2-\frac{\Gamma^2}{4}\right)^{\frac12},
    \end{equation} 
    where $\Gamma=\gamma/m$ denotes the damping rate $\gamma$ per unit mass. 
    This regime is illustrated in the top row in \cref{fig:evolution}, where the short WPP pulse gives a small extension and a negligible contribution of the quartic term in the quasi-parabolic potential. 

\subsection{Time evolution in quasi-parabolic potential}
\label{sec:arg}
    In the case of a quasi-parabolic trapping potential, an additional cubic term $\kappa_3$ appears in the force profile, and it induces a quartic nonlinearity in the quasi-parabolic potential:
    \begin{eqnarray}
        F(z) &=& -\kappa z + \kappa_3 z^3 = -\kappa z (1-\xi z^2), \label{eq:force} \\
        U(z) &=& \frac{1}{2} \kappa z^2 \left( 1-\frac{1}{2} \xi z^2 \right)  \label{eq:potential} \\
        \xi&=&\frac{\kappa_3}{\kappa},  \label{eq:xi}
    \end{eqnarray}
    The oscillator evolving under this type of restoring force is called the (undamped) Duffing oscillator, and for weak Duffing nonlinearity $\xi$, its oscillation frequency $\Omega$ depends on the amplitude of the oscillations $z_\mathrm{max}$ \cite{strogatz_nonlinear_2007,kovacic_duffing_2011}  
    
    \begin{equation}
     \Omega \simeq \Omega_0 \left(1 - \frac38\xi z_\mathrm{max}^2\right). 
     \label{eq:dufomega}
    \end{equation} 
    
    As implied by \cref{eq:dufomega}, the evolution of the state in a quasi-parabolic trapping potential involves a non-uniform rotation with the frequency $\Omega$ that depends on the square of the distance of the point in the phase space from the phase space origin. 
    Consequently, it causes a nonlinear transformation of the state that becomes non-Gaussian.
    This behaviour is shown in the bottom row of \cref{fig:evolution} where 
    the tails of the state extend to the nonlinear region of the quasi-parabolic potential, angularly lag behind the central part of the PDF, and spiral-like arms of the PDF are formed at longer times. 
    In this case, the time evolution of the phase of the radius vector in phase space can be assumed as a rotation with angular frequency $\omega$, as in the parabolic potential, with an additional nonlinear phase shift $\Delta \varphi$: 
    \begin{equation}
        \varphi(t) = - \omega \left(1 - \frac38\bar \xi \bar r^2\right) t + \Phi_0 = -\omega t + \Delta \varphi,
        \label{eq:varphi} 
    \end{equation}
    where $\bar r^2 = \bar z^2 + \bar v^2$ and $\Phi_0$ is the initial phase and 
    \begin{eqnarray}
        \Delta \varphi &=&  \frac38\bar \xi  \omega t \bar r^2 + \Phi_0 \equiv a(t) \bar r^2 + \Phi_0, \\
        \bar \xi &=& \frac{\xi k_\mr{B}T}{m\Omega_0^2}. \label{eq:barxi}
    \end{eqnarray}
    For $\Phi_0=0$, $\Delta \varphi$ directly describes Fermat's spiral with symmetric parabolic arms evolving around the coordinate center $\bar r=\pm \sqrt{\Delta \varphi/a}$ in the two-dimensional phase space. 
    In this case, Fermat's spiral \cite{lockwood_book_2007} is characterized at each time by the scaling parameter $a(t)$ 
    \begin{equation}
        a(t) = \frac38\bar \xi  \omega t.
        \label{eq:FSscale}
    \end{equation}

\section{Transformations of the phase space coordinate system}
    As argued in the previous section, the optimal coordinate transformation should convert an originally non-Gaussian state described in the physical coordinates ($\bar{z},\bar{v}$) to a Gaussian state with no correlation between the transformed coordinates.
    We compare three transformations of the phase space coordinate systems; the linear one is usually used for Gaussian states and is based on eigenvectors of the covariance matrix, and the other two are nonlinear and described below.  
    These transformations are applied to the PDF expressed in physical coordinates, and in each transformed coordinate system  
    the standard deviations $\bar{\sigma}_\mr{ext}$, $\bar{\sigma}_\mr{sqz}$ are evaluated,
    where $\bar{\sigma}_\mr{ext}$ corresponds to the direction of the maximal extension and $\bar{\sigma}_\mr{sqz}$ corresponds to the direction of the maximal squeezing. 
    For comparison, the maximal ($\bar \sigma_{\mathrm{ext,\,HO}}$) and minimal ($\bar \sigma_{\mathrm{sqz,\,HO}}$) standard deviations corresponding to the ideal harmonic oscillator were calculated from Eq. (\ref{eq:Thetadef}) \cite{duchan_nanomechanical_2025}.
    Due to the normalization, all these quantities directly express the extension or squeezing factors with respect to the thermal state. 
    
\begin{figure}[H]
        \centering
        \includegraphics[width=0.5\linewidth]{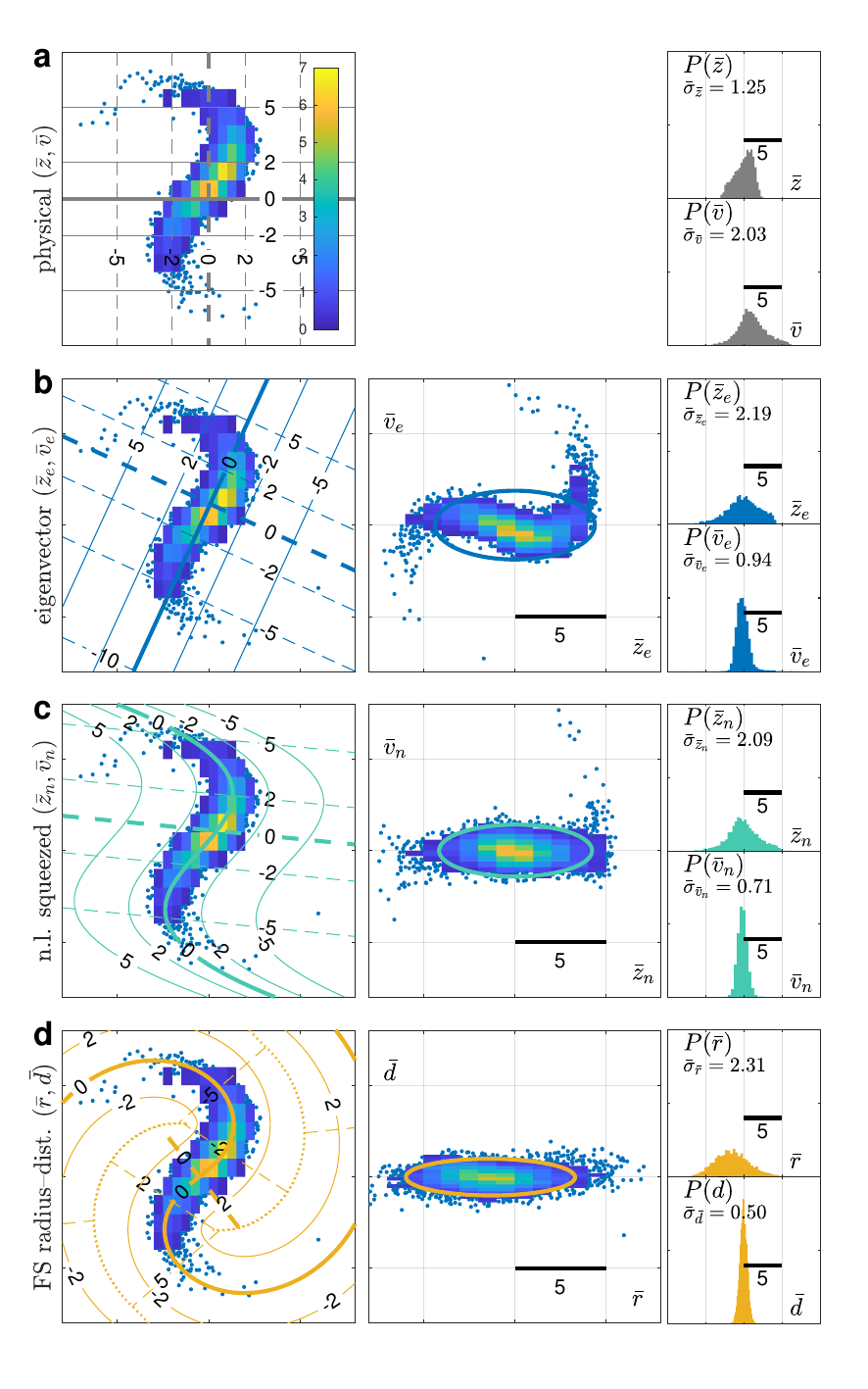}
        \caption{\textbf{Illustration of various coordinate systems used to transform the two-dimensional non-Gaussian PDF from the physical coordinate system to a new system of coordinates to obtain a Gaussian PDF.}
        Colored maps show the probability (in percent) of detecting a particle in a given element of phase space before (left column) and after the transformation (middle column). Individual dots show less-frequently detected events. The solid and dashed lines/curves in the left column depict the local (curvilinear) coordinate system, with thicker curves showing the corresponding coordinate axes. The yellow dotted curve in \textbf{d} indicates identical distance $\bar d$ to both Fermat's spiral arms and consequently the discontinuity in $\bar r$ coordinate. The panels in the right column display the probability density of extended (top) and squeezed (bottom) phase space locations in curvilinear coordinates, along with the corresponding values of the standard deviations. Due to the normalization, in all considered coordinate systems, the standard deviations of the position and velocity of an ideal harmonic oscillator in the state of thermal equilibrium are identically equal to 1. The data correspond to time $t = \SI{14.75}{\micro\second}$ after the IPP pulse of duration $\tau = \SI{2}{\micro\second}$. The whole time evolution of the state in various coordinate systems is available in the attached movie.}
        \label{fig:coords}
    \end{figure}

    Each row of \cref{fig:coords} illustrates the original two-dimensional PDF at a particular time with corresponding transformed system of coordinates (left column), original PDF transformed to such a system of coordinates where extension direction follows the horizontal axis (middle column), and probability density functions along transformed axis (right column) with corresponding values of $\bar{\sigma}_\mr{ext}$, $\bar{\sigma}_\mr{sqz}$. The description of each transformation is provided below and further details can be found in SI.

\subsection{Linear coordinates based on eigenvectors of the covariance matrix: ($\bar z_\mathrm{e}$, $\bar v_\mathrm{e}$) } 
    For a state defined by its mean values and covariance matrix (as shown in \cref{eq:GaussPDF,eq:GaussCM}), a linear transformation can be applied to move from the physical coordinate system shown in \cref{fig:coords}a to a quasi-rotating reference frame with decorrelated axes shown in \cref{fig:coords}b at each time. The following equations give the transformation:
    \begin{align}
    \bar z_\mathrm{e}  &= \phantom{-} (\bar z - \bar z_0) \cos \Phi_\mathrm{e}  + (\bar v - \bar v_0) \sin \Phi_\mathrm{e}  \label{eq:eigenz}\\
    \bar v_\mathrm{e}  &= -(\bar z - \bar z_0) \sin \Phi_\mathrm{e}  + (\bar v - \bar v_0) \cos \Phi_\mathrm{e}. \label{eq:eigenv}
    \end{align}
    Here, $\bar z_0$ and $\bar v_0$ are generally time-dependent mean values of the particle state, and $\Phi_\mathrm{e}$ is the time-varying angle between the $\bar z$ axis and the longer eigenvector of the covariance matrix with all quantities expressed in the physical coordinate system. In the case of Gaussian states uniformly rotating in the physical coordinate system, the state extension and squeezing would correspond to the length of the major and minor semi-axes. The ellipse drawn along the PDF depicts the distance of two standard deviations (along principal axes). After the decorrelation of the coordinate system, these lengths also correspond to the standard deviations $\bar \sigma_{\mr{ext}}$, respectively $\bar \sigma_{\mr{sqz}}$ in the new coordinate system. The coordinate $\bar z_\mathrm{e}$ corresponds to the direction where PDF is extended and $\bar v_\mathrm{e}$ to the direction of squeezing, see \cref{fig:coords}b.

\subsection{Nonlinear squeezed coordinates in a quasi-rotating frame: ($\bar z_\mathrm{n}$, $\bar v_\mathrm{n}$)} 
    This approach was inspired by Ornigotti et al. \cite{ornigotti_nonlinear_2024} and extended to a nonlinearity with higher-order polynomials. Nonlinear squeezing is characterized by a coordinate transformation that subtracts higher powers of the extended transformed coordinate $\bar{z}_\mr{n}$ from the squeezed one $\bar{v}_\mr{n}$. Such higher-order polynomial variables can be consistently measured and hierarchically analyzed also down to the quantum regime \cite{moore_estimation_2019, moore_hierarchy_2022}. We change the coordinate system to a quasi-rotating frame of reference, as in the previous case, and add the suggested nonlinear term. The coordinate transformation can be expressed as:
    \begin{align}
        \bar z_\mathrm{n} & = \phantom{-}(\bar z - \bar  z_0) \cos \Phi_\mathrm{n} + ( \bar v - \bar v_0)\sin \Phi_\mathrm{n} , \label{eq:nlinz} \\
        \bar v_\mathrm{n} & = -(\bar z - \bar  z_0)\sin \Phi_\mathrm{n}  + ( \bar v - \bar v_0)\cos \Phi_\mathrm{n}  -  \sum_{k=1}^N \lambda_k \bar z_\mathrm{n}^k,
        \label{eq:nlin}
    \end{align}
    where $\Phi_\mathrm{n}$ is the angle of the rotating frame and $\lambda_k$ are unknown coefficients representing the contribution of the nonlinear state deformation by higher order polynomials. Since a particle's squeezed width has a positive lower bound, one can search for the optimal $\lambda_k$ values that minimize the variance of $\bar v_\mathrm{n}$, see \cite{ornigotti_nonlinear_2024}.
    
    Compared to the eigenvectors-based coordinates in \cref{fig:coords}b, the major coordinate axis of the nonlinear squeezed coordinates in \cref{fig:coords}c (green thick solid line) follows better the shape of the particle state, including the quartic nonlinearity-induced twist. 
    This transformation successfully narrows the distribution in the squeezed variable, see the right panel in \cref{fig:coords}c). However, the extended variable also appears slightly narrower, and the tips of the state tails (individually scattered points in the middle panel of \cref{fig:coords}c) are not correctly aligned with the new coordinate axis. That indicates the method doesn't accurately capture the state's full extent.

\subsection{Nonlinear coordinates based on Fermat's spiral: ($\bar r$, $\bar d$)}
    We propose here a nonlinear coordinate transformation based on Fermat's spiral to accurately capture the non-Gaussian shape of measured classical stochastic states of a nanoparticle levitated in a quasi-parabolic optical trapping potential with a weak quartic term. At each measured time of the nanoparticle state evolution, we look for the shortest length of Fermat's spiral that gives the minimal sum of distances of phase space points from it. We start the search with an arbitrary Fermat's spiral centered at \((\bar z_0, \bar v_0)\) characterized by an instantaneous scaling factor $a$ and angle $\Psi_\mr{r}$ describing the instantaneous rotation of the new coordinate system with respect to the fixed physical coordinate system, where the total spiral angle is \(\phi = a \bar r^2 + \Phi_\mathrm{r}\). We define the Euclidean distance between any phase space point $(\bar{z},\bar{v})$ and the spiral as:
    \begin{equation}
        \bar d = \min \left[ \left(\bar z - \bar z_0 - \bar r\cos \phi \right)^2 + \left(\bar v - \bar v_0 - \bar r\sin \phi  \right)^2   \right]^{\frac12}
    \end{equation}
    Here, \(\bar d\) is the squeezed coordinate, and $\bar r$ is the complementary extended coordinate. All experimental points are transformed from the physical coordinates \((\bar z, \bar v)\) to Fermat's spiral coordinates \((\bar r, \bar d)\), where minimization in $\bar d$ is performed numerically for each experimental point in the phase space at given time~$t$. In the new coordinate system \((\bar r, \bar d)\),  the four spiral parameters \((\bar z_0, \bar v_0, a, \Phi_\mathrm{r})\) are further optimized by minimizing the variance of the squeezed variable \(\bar d\) with a regularization term accounting for the total length $L$ of the spiral arms:
    \begin{equation}
        (\bar z_0, \bar v_0, a, \Phi_\mathrm{r}) = \mathrm{argmin} \left[\mathrm{var\ } \bar d + \lambda_\mathrm{R} L \right].
    \end{equation}
    We found an optimal regularization strength for our system of \(10^{-3} \leq \lambda_\mathrm{R} \leq 10^{-2}\).
    
    Figure \ref{fig:coords}d shows that the major coordinate axis $\bar{r}$ (thick solid yellow line) follows well both the central part of the particle state as well as the sparsely populated tails. 
    The side effect is the discontinuity of $\bar r$ depicted by the yellow dotted curve.
    The right panels of \cref{fig:coords}d show even further state squeezing in $\bar d$ (yellow bottom histogram) to the value of almost half of the squeezing obtained in the eigenvector coordinates. 
    In this coordinate system, the state is also extended in $\bar r$.

\section{Results and discussion}
\begin{figure}[H]
        \centering
        \includegraphics[width=1\linewidth]{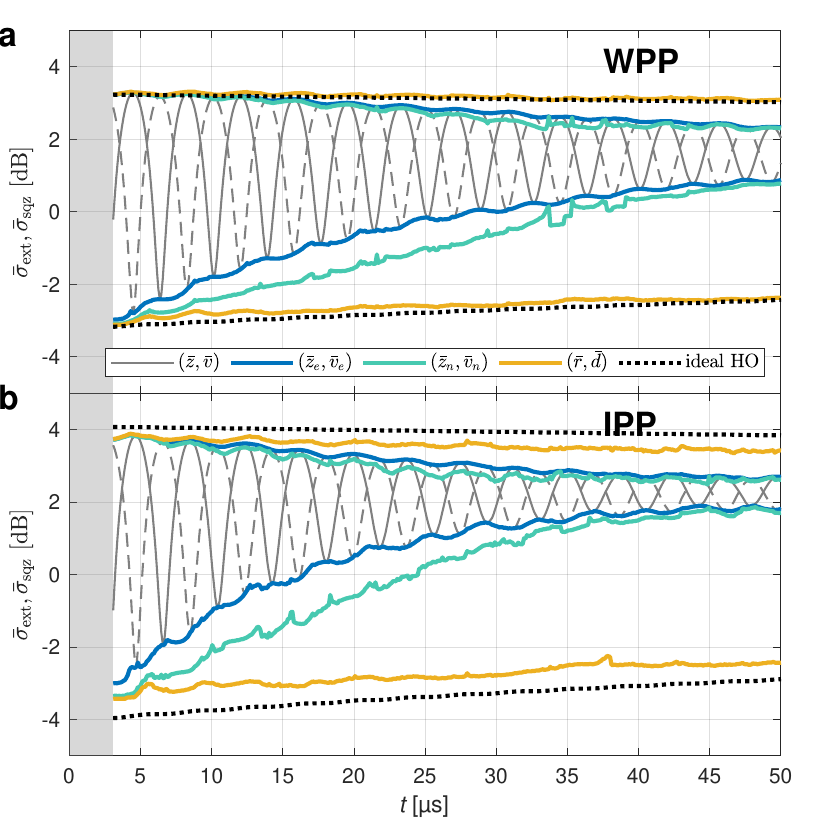}
        \caption{\textbf{Comparison of time evolution of the state extension and squeezing factors $\bar{\sigma}_\mr{ext}$ and $\bar{\sigma}_\mr{sqz}$ at different coordinate systems.} 
        Solid and dashed gray curves correspond to the standard deviations of physical coordinates $\bar z$ and $\bar v$, respectively. Blue curves represent the extension and squeezing factors obtained in linear coordinates $\bar z_\mr{e}$ and $\bar v_\mr{e}$ based on eigenvectors of the covariance matrix. The green and yellow curves correspond to the nonlinear squeezed coordinates $\bar z_\mr{n}$, $\bar v_\mr{n}$ in a quasi-rotating frame and the nonlinear coordinates based on Fermat's spiral $\bar r$, $\bar d$. The parameters found for each transformation at the given time are plotted in \cref{fig:coordpar2} and discussed in the SI. 
        Black dotted curves correspond to the ideal theoretical 
        extension and squeezing factors $\bar{\sigma}_\mr{ext,\,HO}$ and $\bar{\sigma}_\mr{sqz,\,HO}$,  
        that represent the ideal limits of the state extension.
        Gray regions correspond to the detector dead-time ( $\approx$ \SI{3}{\micro \s}) when trustworthy data can not be acquired. 
        The duration of WPP and IPP pulses was 
        $\tau = \SI{2}{\micro\second}$. }
        \label{fig:sigmat}
    \end{figure}

\begin{figure}[H]
        \centering
        \includegraphics[width=1\linewidth]{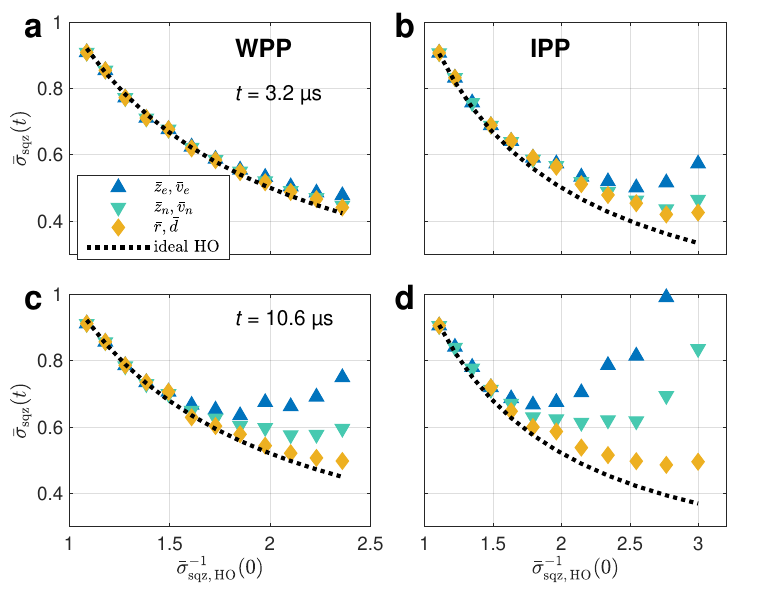}
        \caption{\textbf{Comparison of apparent state squeezing in different coordinate systems.}
        Black dotted curves correspond to the theoretical limits of the maximal squeezing factors of the Gaussian state evolving in a purely parabolic potential ($\bar{\sigma}_\mr{sqz,\,HO}$). 
        Blue triangles represent the extension and squeezing factors obtained in linear coordinates based on eigenvectors of the covariance matrix, the green triangles correspond to the nonlinear squeezed coordinates in a quasi-rotating frame, and the yellow diamonds to the nonlinear coordinates based on Fermat's spiral.}
        \label{fig:squuzecmp}
    \end{figure}

    While the previous session and \cref{fig:coords} analyze the experimental motional state at a particular time, this section will cover its time evolution (\cref{fig:sigmat}) and different initial conditions (\cref{fig:squuzecmp}). 
    The experimental results are compared with the theoretical results for maximal and minimal standard deviations  $\bar \sigma_{\mathrm{ext,\,HO}}$ and $\bar \sigma_{\mathrm{sqz,\,HO}}$ of the damped harmonic oscillator\cite{duchan_nanomechanical_2025}. 

    Considering the experimental input parameters and gradual reheating due to contact of the levitated nanoparticle with the ambient thermal environment, they represent the ideal performance limit of the studied coordinate transformations.
    
    Following \cref{eq:eigenz,eq:eigenv}, the blue curves in  \cref{fig:sigmat} form the envelopes of the oscillating gray curves of physical standard deviations of $\bar z$ and $\bar v$. 
    This transformation removes linear correlations between the position and velocity in the physical coordinate system, but it does not address those arising from nonlinearity. This is evident from the difference between the dotted curves, which are expected for the ideal parabolic potential, and the blue curves. 
    
    The green curves in \cref{fig:sigmat} depict the extension and squeezing factors after the nonlinear coordinates transformation to $\bar z_\mr{n}$ and $\bar v_\mr{n}$ defined in \cref{eq:nlinz,eq:nlin}. This method suppresses the nonlinearity effect for short times but gives similar results as the previous linear method for longer times.  
    
    The yellow curves in \cref{fig:sigmat} employ the transformation based on Fermat's spiral ($\bar r, \bar d$) and represent the best results we obtained. 
    In the case of WPP, the squeezing and extension factors are almost identical to the values obtained for the ideal parabolic potential (dotted), including the interaction with the gas molecules remaining in the vacuum chamber, which leads to the so-called reheating effect. The similar average slope of the ideal theoretical limit and the transformed experimental data (positive slope for $\bar{\sigma}_\mr{sqz}$ and negative slope for $\bar{\sigma}_\mr{ext}$)  indicates that the nonlinearity was separated from the reheating effect.
    
    For the case of IPP, the extension and squeezing results obtained for Fermat's spiral coordinate system and the ideal model of the damped harmonic oscillator (dotted) slightly deviate. 
    This is most likely due to the displacement of the experimental IPP with respect to the quasi-parabolic potential. It causes a slight asymmetry between both arms of Fermat's spirals, which are, however, considered as identical during the coordinate transformation. 

    \Cref{fig:squuzecmp} compares the squeezing factor $\bar{\sigma}_\mr{sqz}$ (top row) of the initial state and a state approximately one oscillation period later (bottom row), evaluated for each transformation of the coordinate systems described above. The squeezing factor of the initial state depends on the duration $\tau$ of the WPP or IPP potential pulse, as illustrated in the SI. 
    
    Regarding the initial squeezing factors for the WPP and all coordinate systems, they are all close to each other and also to the expected theoretical limit obtained for the harmonic potential (dotted). 
    However, the initial squeezing factors obtained using the IPP start to differ from the harmonic potential limit for $\tau > \SI{1.3}{\micro\second}$.
    Due to the displaced IPP with respect to the quasi-parabolic potential, the center of the PDF is shifted with respect to the center of the physical coordinate system, and both spiral arms evolve in time differently. Such generated asymmetry is not part of the nonlinear transformations or the theoretical model, and it leads to larger discrepancies between the results.  
    
    The second row in \cref{fig:squuzecmp} compares the squeezing factors evaluated in the quasi-parabolic potential, approximately one oscillation period later. By this time, the motional state evolved towards Fermat's spiral and significant deviations arose between all transformations. As expected, the squeezing factor obtained in the eigenvector coordinate system (blue triangles) significantly deviates from the results of the nonlinear transformations and the harmonic oscillator. On the other hand, Fermat's spiral-based coordinate system characterizes the squeezing very well even for long-duration pulses, especially in the case of symmetric motional states generated by the centered WPP potential pulses.

\section{Conclusion}
    We analyzed the experimentally recorded stochastic motion of a nanoparticle optically levitated in 
    a standing wave optical trap at 
    a gaseous thermal environment of pressure 1 mbar. 
    By rapidly switching the optical standing wave, we generated a potential pulse in the form of a weak parabolic potential or an inverted parabolic potential, thereby setting the nanoparticle out of thermal equilibrium.  We detected its subsequent time evolution in the standing wave optical trap, corresponding to a quasi-parabolic potential with a weak quartic (Duffing) term. We acquired nanoparticle trajectories for 5000 repetitions of the protocol.
    Analyzing the data, we determined the two-dimensional probability density function in the phase space at each measured time. 
    
    The potential pulse led to extension/squeezing and displacement of the probability density function. In the case of larger extensions, it brought the nanoparticle outside the region of harmonic confinement. It reached a nonlinear region of the quasi-parabolic optical potential, which is approximated here as a Duffing-type quartic nonlinearity. It led to nonlinear distortions of the original Gaussian probability density function. As a result, the mean values and the standard deviations or variances of the state
    along the characteristic directions of the state extension and squeezing no longer accurately describe the instantaneous state and its temporal evolution. 
    
    To decouple the nonlinearity of the nanoparticle's motional dynamics from the linear extension/squeezing 
    and the gradual thermalization of the nanoparticle with the ambient gas, we proposed a nonlinear transformation of the phase space coordinate system that effectively restores the Gaussian character of the nanoparticle's state. 
    
    Specifically, for the optical trapping potential with a Duffing-type quartic term, the coordinate transformation derived from Fermat's spiral geometry was adopted. 
    We systematically studied the temporal evolution of the variance-based state characteristics (extension and squeezing factors) in the transformed coordinates following a wide range of initial Gaussian states. 
    
    We compared the experimental data with the ideal theoretical limit of a Gaussian state evolving in a pure parabolic potential. This comparison showed that Fermat's spiral coordinates, which accurately reflect the dynamics of the actual studied nonlinear system, allow the formal description of such a system using the language of variances of Gaussian states, typically limited to linear systems. 
    
    Fermat's spiral-based coordinates represent an analogy to the rotating frame coordinates used in the linear regime. While the rotating frame is governed by the phase that linearly grows in time, the nonlinear Fermat's spiral-based coordinates add a parameter -- scale $a(t)$ that grows linearly with time as well.
    Our results can be applied in devising advanced protocols for levitated nonlinear optomechanics, including non-Gaussian states.

\begin{acknowledgement}
We acknowledge support by the projects CZ.02.01.01/00/22\_008/0004649 of M\v{S}MT \v{C}R and GA23-06224S of GA \v{C}R.
\end{acknowledgement}

\begin{suppinfo}

\section*{Experimental details}
\label{sec:exp}
    The full experimental setup and protocols were described in detail in Ref. \cite{duchan_nanomechanical_2025} and therefore only a summary is presented here. 
    We levitated a silica nanoparticle of radius $a\approx \SI{150}{\nm}$ in a standing wave trap formed by two counter-propagating Gaussian laser beams of wavelength $\SI{1064}{\nm}$. The beams, each of power 20 mW,  were focused into a vacuum chamber to a Gaussian beam waist of radius  $ \approx~\SI{1}{\micro\m}$.  The nanoparticle is levitated in the vicinity of the beam foci at ambient pressure \SI{1}{\mbar}. 
    The effects studied in this paper are limited to nanoparticle motion along the optical axis ($z$-axis). The mechanical oscillation frequency of the nanoparticle  $\omega/2\pi\approx~\SI{135}{\kHz}$ was obtained by analyzing the power spectral density of the levitated nanoparticle \cite{hebestreit_calibration_2018}.
    
    The initial Gaussian state, defined by its mean values $\vmzi$ and covariance matrix  $\mathbf \Theta_0$ at $t=0$ was formed by a sequence of potential pulses as illustrated in \cref{fig:protocol} and described in Ref. \cite{duchan_nanomechanical_2025}. While the minima of the WPP (\cref{fig:protocol}a) and quasi-parabolic potential overlap, the maximum of the IPP and the minimum of the quasi-parabolic potential are displaced by $\Delta \approx$ 70~nm (\cref{fig:protocol}b). Such displacement is equivalent to an additional constant force $F_\mathrm{c} = \kappa_\mathrm{i}\Delta$ acting upon the nanoparticle during the pulse, where $\kappa_\mathrm{i}$ denotes the stiffness of the inverted potential. Such force displaces the motional state in the phase space, i.e. it shifts the mean values of the PDF. 
    
    The particle is exposed to the WPP or IPP for a time $\tau$ (see \cref{fig:protocol}c) which determines the extension/squeezing factor of the initial Gaussian state.
    In this paper, the initial state at time $t=\SI{0}{\micro\s}$ corresponds to the moment right after the pulse of WPP or IPP.
    We correctly detect the nanoparticle position in the quasi-parabolic potential after the pulse and recovery of the detection system ($\approx \SI{3}{\micro\second}$).

    The WPP and IPP protocols are repeated independently multiple times (5000 each)  while the nanoparticle position in the quasi-parabolic potential is detected with a sampling frequency  $\approx \SI{10}{\MHz}$. 
    The particle velocity is then calculated from the low-pass filtered position data as the second-order central difference of the positions. 
    This way, statistical ensembles are collected, allowing us to reconstruct the particle motional state \cite{duchan_nanomechanical_2025}. 
    
    \begin{figure}[H]
        \centering
        \includegraphics[width=1\linewidth]{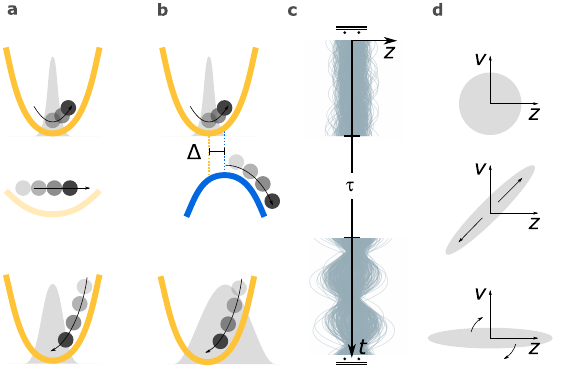}
        \caption{\textbf{Preparation of the initial state in the quasi-parabolic potential.} \textbf{a} A nanoparticle levitates in parabolic potential (top), which is switched to a weak quasi-parabolic potential (WPP) for time $\tau$ and then switched to a quasi-parabolic potential (bottom). \textbf{b} The same as \textbf{a} except switching to an inverted parabolic potential (IPP), its maximum is displaced with respect to the minimum of the parabolic potential by $\Delta$. \textbf{c} Overlapping trajectories in the z-axis acquired after repetition of the steps a) or b) 5000 times. Corresponding probability density functions are illustrated as gray silhouettes in \textbf{a,b}. \textbf{d} The corresponding motional states, illustrated as the shape of two-dimensional probability density functions (PDF) in the phase space, in the parabolic potential (top), after the potential pulse causing state extension/squeezing (WPP, IPP) and displacement (IPP) (middle) and rotated and deformed state in the quasi-parabolic potential (bottom).}
        \label{fig:protocol}
    \end{figure}

\section*{Determined parameters of coordinate systems}
\label{sec:trans}
    \Cref{fig:coordpar2} plots the time evolution of the found parameters for each coordinate transformation and both WPP (left column) and IPP (right column) pulses of duration $\tau = \SI{2}{\micro\second}$. 
    Figures \ref{fig:coordpar2}a,b show the evolution of the ``reference point'' or offset of the coordinate transform, i.e. optimal quantities $\bar z_0$ and $\bar v_0$.
    In the case of the covariance matrix eigenvectors $\bar z_\mr{e}$ and $\bar v_\mr{e}$, $\bar z_0$ and $\bar v_0$ correspond by definition to the mean values of the state in the physical coordinates $\bar z$ and $\bar v$.
    In the case of nonlinear squeezed coordinates in a quasi-rotating frame, $\bar z_\mr{n}$ and $\bar v_\mr{n}$, the green curves for $\bar z_0$ and $\bar v_0$ follow tightly 
    the previously discussed mean values (blue). 
    On the other hand, Fermat's spiral-based coordinate system (yellow) provides optimized values of $\bar{z}_0$ and $\bar{v}_0$ that are approximately one-half of the previously discussed transformations, and their oscillation is phase-shifted by approximately 60 degrees behind the oscillations of the other methods. 
    It results from the fact that the position of the ``reference point'' for Fermat's spiral-based coordinate system follows more closely the motion of the most probable value of the PDF (particle motional state) rather than its mean value, which are shifted from each other due to the slight state asymmetry. 
    The displacement of the IPP with respect to the quasi-parabolic potential leads to the displacement of the mean values from the center of the physical coordinate system \cite{duchan_nanomechanical_2025}, which is demonstrated by an order of magnitude larger magnitude of the oscillations of found $\bar z_0$ and $\bar v_0$ in contrast to the WPP results. Slight damping of these oscillations is also visible.  
    
    Figures \ref{fig:coordpar2}c show the ratio of the rotational frequency of the state, determined from the coordinate rotation angle $\Phi_{e,n,r}$ at each time (smoothened by the second order Savitzky-Golay filter with 15 points window), and the oscillation frequency of the nanoparticle in the optical trap obtained from the power spectral density fitting \cite{hebestreit_calibration_2018}. The results obtained in this manner are close to the expected value of 1.
    
    Figures \ref{fig:coordpar2}d show the additional three parameters $\lambda_1,\,\lambda_2,\,\lambda_3$ that were found to optimize the nonlinear transformation to coordinates $\bar z_\mr{n}$ and $\bar v_\mr{n}$.  
    The linear term characterized by $\lambda_1$ is responsible for the skewed angle at the intersection of coordinate axes in the ``reference point,'' see thick solid and dashed curves in \cref{fig:coordpar2}c. 
    
    Values of $\lambda_1$ reach the extreme near $t=\SI{25}{\micro\second}$ and then gradually go to zero while becoming more unstable. This indicates that this curvilinear system no longer accurately describes the state and shifts closer to the eigenvector-based description. 
    
    The term $\lambda_{3}$ is connected to the Duffing nonlinearity. It linearly increases with time to $t\sim \SI{25}{\micro\second}$, indicating the stronger nonlinear term, but later goes to zero and becomes unstable similarly to $\lambda_1$.
    Strong term $\lambda_2$ in the IPP case is induced by the motion of the displaced state in the quasi-parabolic potential, where nonlinearity causes the state asymmetry. 
    In contrast, this term is negligible for the WPP case, where the minima of WPP and the quasi-parabolic potential overlap. 
    
    \Cref{fig:coordpar2}e plots the scaling parameter $a(t)$ of the optimized Fermat's spiral transformation which follows \cref{eq:FSscale} and linearly grows in time. 
    Assuming the nanoparticle motion in the standing wave, the restoring optical force near the potential minimum $F\simeq -\sin(2 k z) \simeq -2 k z [1-(2 k z)^2/6]$. Using Eq. (\ref{eq:force}) it gives $\xi=(2 k)^2/6$ and following Eq. (\ref{eq:barxi}) one gets 
    \begin{equation}
    \bar \xi = \frac23\left(\frac{2\pi}{\lambda}\right)^2 \frac{k_\mr{B}T}{m\Omega_0^2}.  \label{eq:barxisw}  
    \end{equation} 
    Using the experimental parameters, the solid blue line shows the value of the scaling parameter calculated using \cref{eq:FSscale}. Since the mass (radius) of the particle is not known precisely, the dashed line corresponds to the scaling parameter obtained for the particle radius smaller by 10\% than the expected value of \SI{150}{\nm}. Its coincidence with the WPP result is persuasive. 

    \begin{figure}[H]
        \centering
        \includegraphics[width=.8\linewidth]{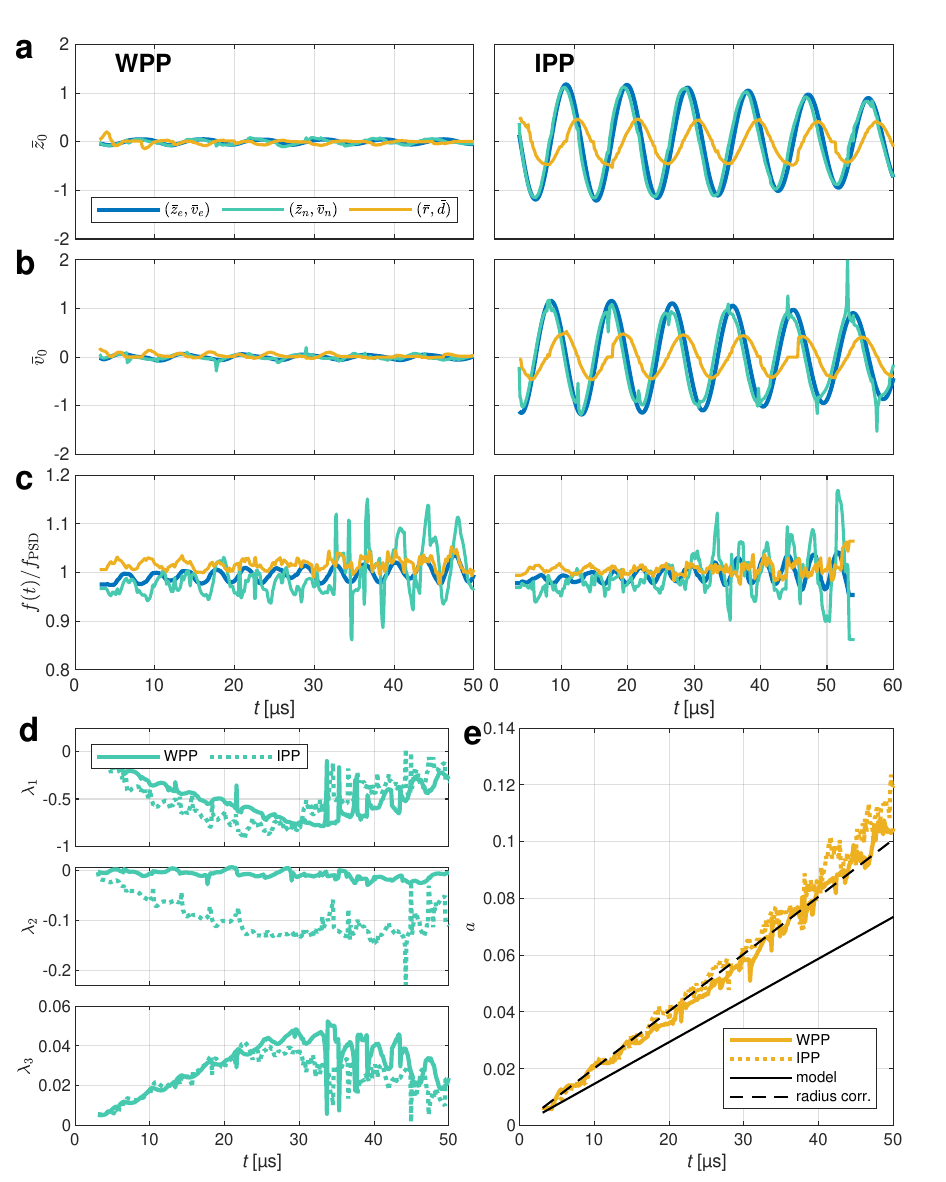}
        \caption{\textbf{Comparison of parameter coordinate systems transformation in WPP and IPP. $\tau = \SI{2}{\micro\s}$}  
        Evolution of the state mean position \textbf{a} and mean velocity \textbf{b} or Fermat's spiral center for WPP (left) and IPP (right) pulse. 
        Each color of the curves represents $\bar z_0$ and $\bar v_0$ in a different coordinate system.
        \textbf{c} Instant state frequency obtained by the time differential of the angle $\Phi_e$,$\Phi_n$, or $\Phi_r$ expressed relatively with respect to the resonant frequency obtained from the power spectral density $f_\mr{PSD}$.     
        \textbf{d} Time evolution of the polynomial coefficients $\lambda_1,\, \lambda_2,\,\lambda_3$ of nonlinear contributions to the state deformation, see Eq. (\ref{eq:nlin}). Solid and dotted curves correspond to WPP or IPP, respectively. 
        \textbf{e} Evolution of Fermat's spiral scaling parameter $a$ for WPP (solid yellow) and IPP (dotted yellow). 
        Solid and dashed black lines show the evolution of $a$ based on the theoretical estimate of the  Duffing nonlinearity in the optical standing wave (see \cref{eq:barxisw}) for the particle radius \SI{150}{\nm} and \SI{135}{\nm}, respectively.
        }
        \label{fig:coordpar2}
    \end{figure}

\end{suppinfo}

\providecommand{\latin}[1]{#1}
\makeatletter
\providecommand{\doi}
  {\begingroup\let\do\@makeother\dospecials
  \catcode`\{=1 \catcode`\}=2 \doi@aux}
\providecommand{\doi@aux}[1]{\endgroup\texttt{#1}}
\makeatother
\providecommand*\mcitethebibliography{\thebibliography}
\csname @ifundefined\endcsname{endmcitethebibliography}
  {\let\endmcitethebibliography\endthebibliography}{}

\end{document}